\documentclass[12pt]{article}

\usepackage[utf8]{inputenc}
\usepackage{color}
\usepackage{url}
\usepackage{graphicx}
\usepackage[margin=1in]{geometry}

\begin{document}

\title{Recent Results on\\Classifying Risk-Based Testing Approaches}
\author{Michael Felderer \\ \emph{University of Innsbruck, Austria} \vspace{.5cm} \\ J\"urgen Gro\ss{}mann \\ \emph{Fraunhofer FOKUS, Germany} \vspace{.5cm} \\ Ina Schieferdecker \\ \emph{Fraunhofer FOKUS \& TU Berlin, Germany}}
\date{}
\maketitle

\abstract{In order to optimize the usage of testing efforts and to assess risks of software-based systems, risk-based testing uses risk (re-)assessments to steer all phases in a test process. Several risk-based testing approaches have been proposed in academia and/or applied in industry, so that the determination of principal concepts and methods in risk-based testing is needed to enable a comparison of the weaknesses and strengths of different risk-based testing approaches. In this chapter we provide an (updated) taxonomy of risk-based testing aligned with risk considerations in all phases of a test process. It consists of three top-level classes, i.e., contextual setup, risk assessment, and risk-based test strategy. This taxonomy provides a framework to understand, categorize, assess and compare risk-based testing approaches to support their selection and tailoring for specific purposes. Furthermore, we position four recent risk-based testing approaches into the taxonomy in order to demonstrate its application and alignment with available risk-based testing approaches.}

\section{Introduction}
\label{sec:introduction}

Testing of safety-critical, security-critical or mission-critical software faces the problem of determining those tests that assure the essential properties of the software and have the ability to unveil those software failures that harm the critical functions of the software. However, also for "normal" less critical software a comparable problem exists: Usually testing has to be done under severe pressure due to limited resources and tight time constraints with the consequence that testing efforts have to be focused and be driven by business risks. 

Both decision problems can adequately be addressed by risk-based testing which consider risks of the software product as the guiding factor to steer all phases of a test process, i.e., test planning, design, implementation, execution, and evaluation~\cite{gerrard2002risk,felderer2014integrating,felderer2014taxonomy}. Risk-based testing is a pragmatic, in companies of all sizes widely used approach~\cite{felderer2014multiple,felderer2015sme} which uses the straightforward idea to focus test activities on those scenarios that trigger the most critical situations of a software system~\cite{wendland2012systematic}. 

Recently, the international standard ISO/IEC/IEEE 29119 Software Testing~\cite{ISO2013SoftwareTesting} on testing techniques, processes, and documentation even explicitly specifies risk considerations to be an integral part of the test planning process. Because of the growing number of available risk-based testing approaches and its increasing dissemination in industrial test processes~\cite{felderer2014framework}, methodological support to categorize, assess, compare, and select risk-based testing approaches is required. 

In this paper, we present an (updated) taxonomy of risk-based testing that provides a framework for understanding, categorizing, assessing, and comparing risk-based testing approaches and that supports the selection and tailoring of risk-based testing approaches for specific purposes. To demonstrate the application of the taxonomy and its alignment with available risk-based testing approaches, we position four recent risk-based testing approaches, i.e., the RASEN approach~\cite{RASEN_D533}, the SmartTesting approach~\cite{ramler2015process}, risk-based test case prioritization based on the notion of risk exposure~\cite{yoon2011test} as well as risk-based testing of open source software~\cite{yahav2014risk}, in the taxonomy.

A \emph{taxonomy} defines a hierarchy of classes (also referred to as categories, dimensions, criteria or characteristics) to categorize things and concepts. It describes a tree structure whose leaves define concrete values to characterize instances in the taxonomy. The proposed taxonomy is aligned with the consideration of risks in all phases of the test process and consists of the top-level classes \emph{context} (with subclasses risk driver, quality property, and risk item), \emph{risk assessment} (with subclasses factor, estimation technique, scale, and degree of automation), and \emph{risk-based test strategy} (with subclasses risk-based test planning, risk-based test design \& implementation, and risk-based test execution \& evaluation). The taxonomy presented in this chapter extends and refines our previous taxonomy of risk-based testing~\cite{felderer2014taxonomy}. 

The remainder of this chapter is structured as follows. Section~\ref{sec:backgr-softw-test} presents background on software testing and risk management. Section~\ref{sec:taxonomy} introduces the taxonomy of risk-based testing. Section~\ref{sec:class-recent-risk} presents the selected four recent risk-based testing approaches and discusses them in the context of the taxonomy. Finally, Section~\ref{sec:summary} summarizes this chapter.

\section{Background on Software Testing and Risk Management}
\label{sec:backgr-softw-test}

\emph{Software testing}~\cite{istqb2012standardGlossary} is the process consisting of all lifecycle activities, both static and dynamic, concerned with planning, preparation and evaluation of software products and related work products to determine that they satisfy specified requirements, to demonstrate that they are fit for purpose and to detect defects. According to this definition it comprises static activities like reviews but also dynamic activities like classic black or white box testing. The tested software-based system is called \emph{system under test} (SUT). As highlighted before, \emph{risk-based testing} (RBT) is a testing approach which considers risks of the software product as the guiding factor to support decisions in all phases of the test process~\cite{gerrard2002risk,felderer2014integrating,felderer2014taxonomy}. A \emph{risk} is a factor that could result in future negative consequences and is usually expressed by its likelihood and impact~\cite{istqb2012standardGlossary}. In software testing, the \emph{likelihood} is typically determined by the probability that a failure assigned to a risk occurs, and the \emph{impact} is determined by the cost or severity of a failure if it occurs in operation. The resulting \emph{risk value} or \emph{risk exposure} is assigned to a \emph{risk item}. In the context of testing, a risk item is anything of value (i.e., an asset) under test, for instance, a requirement, a component or a fault. 

\begin{figure*}[htb]
\begin{center}
  \includegraphics[width=.25\textwidth]{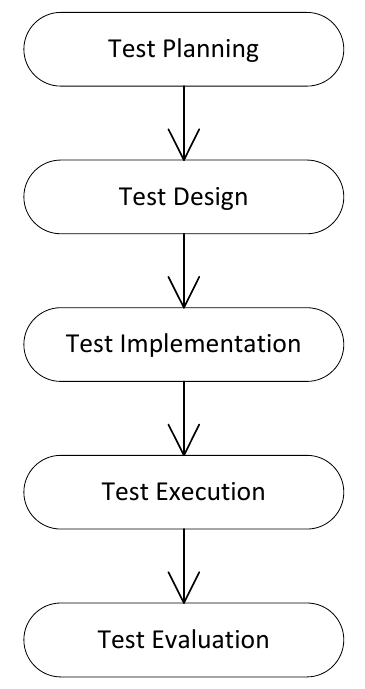}
  \caption{Core test process steps}
  \label{fig:test-process}
\end{center}
\end{figure*}

RBT is a testing-based approach to risk management which can only deliver its full potential if a test process is in place and if risk assessment is integrated appropriately into it. A \emph{test process} consists of the core activities test planning, test design, test implementation, test execution, and test evaluation~\cite{istqb2012standardGlossary} (see Figure~\ref{fig:test-process}). In the following, we explain the particular activities and associated concepts in more detail.

According to~\cite{ISO2013SoftwareTesting} and~\cite{istqb2012standardGlossary}, \emph{test planning} is the activity of establishing or updating a test plan. A test plan is a document describing the scope, approach, resources, and schedule of intended test activities. It identifies, amongst others, objectives, the features to be tested, the test design techniques, and exit criteria to be used and the rationale of their choice. \emph{Test objectives} are reason or purpose for designing and executing a test. The reason is either to check the functional behavior of the system or its non-functional properties. \emph{Functional testing} is concerned with assessing the functional behavior of an SUT, whereas \emph{non-functional testing} aims at assessing non-functional requirements such as security, safety, reliability or performance. The scope of the features to be tested can be components, integration or system. At the scope of \emph{component testing} (also referred to as unit testing), the smallest testable component, e.g., a class, is tested in isolation. \emph{Integration testing} combines components with each other and tests those as a subsystem, that is, not yet a complete system. In \emph{system testing}, the complete system, including all subsystems, is tested. \emph{Regression testing} is the selective retesting of a system or its components to verify that modifications have not caused unintended effects and that the system or the components still comply with the specified requirements~\cite{Radatz1990IEEEStandardGlossary}. \emph{Exit criteria} are conditions for permitting a process to be officially completed. They are used to report against and to plan when to stop testing. Coverage criteria aligned with the tested feature types and the applied test design techniques are typical exit criteria. Once the test plan has been established, test control begins. It is an ongoing activity in which the actual progress is compared against the plan which often results in concrete measures. 

During the \emph{test design} phase the general testing objectives defined in the test plan are transformed into tangible test conditions and abstract test cases. 
\emph{Test implementation} comprises tasks to make the abstract test cases executable. This includes tasks like preparing test harnesses and test data, providing logging support or writing test scripts which are necessary to enable the automated execution of test cases. In the \emph{test execution} phase, the test cases are then executed and all relevant details of the execution are logged and monitored. Finally, in the \emph{test evaluation} phase the exit criteria are evaluated and the logged test results are summarized in a test report. \\

\emph{Risk management} comprises the core activities \emph{risk identification}, \emph{risk analysis}, \emph{risk treatment}, and \emph{risk monitoring}~\cite{ASNZS2004RiskManagement,ISO31000RiskManaagement}. In the risk identification phase, risk items are identified. In the risk analysis phase, the likelihood and impact of risk items and, hence, the risk exposure is estimated. Based on the risk exposure values, the risk items may be prioritized and assigned to risk levels defining a risk classification. In the risk treatment phase the actions for obtaining a satisfactory situation are determined and implemented. In the risk monitoring phase the risks are tracked over time and their status is reported. In addition, the effect of the implemented actions is determined. The activities risk identification and risk analysis are often collectively referred to as \emph{risk assessment}, while the activities risk treatment and risk monitoring are referred to as \emph{risk control}.  

\section{Taxonomy of risk-based testing} 
\label{sec:taxonomy}

\begin{figure*}
  \includegraphics[width=1\textwidth]{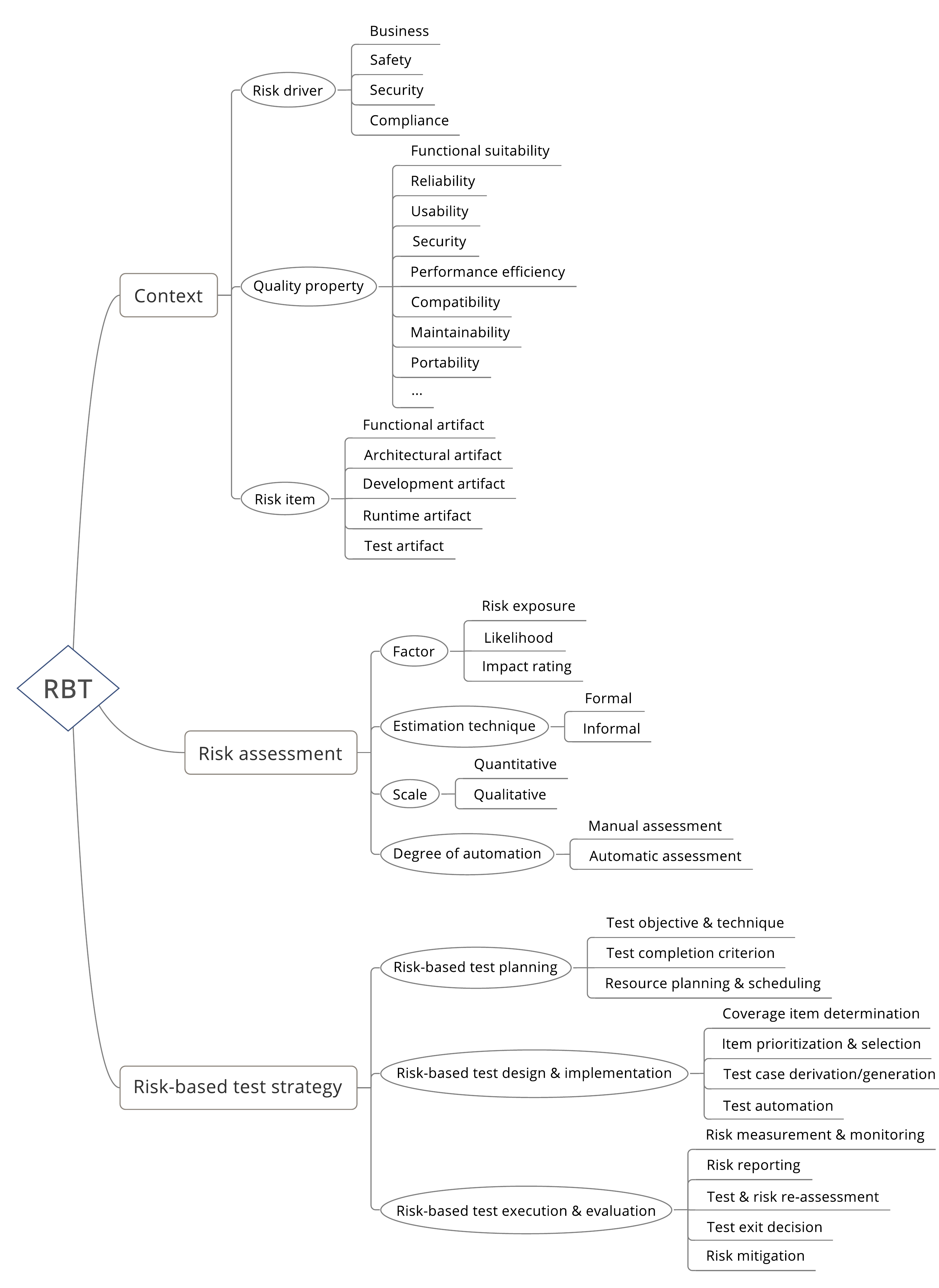}
  \caption{Risk-based testing taxonomy}
  \label{fig:rbt-taxonomy}
\end{figure*}

The taxonomy of risk-based testing is shown in Figure~\ref{fig:rbt-taxonomy}. It contains the top-level classes \emph{contextual set up}, \emph{risk assessment} as well as \emph{risk-based test process} and is aligned with the consideration of risks in all phases of the test process. In this section, we explain these classes, their subclasses and concrete values for each class of the risk-based testing taxonomy in depth. 

\subsection{Context}
\label{sec:contextual-set-up}
The \emph{context} characterizes the overall context of the risk assessment and testing processes. It includes the subclasses \emph{risk driver}, \emph{quality property} and \emph{risk item} to characterize the drivers that determine the major assets, the overall quality objectives that need to be fulfilled and the items that are subject to evaluation by risk assessment and testing.  

 \subsubsection{Risk driver} 
\label{sec:risk-driver}
A \emph{risk driver} is the first differentiating element of risk-based testing approaches. It characterizes the area of origin for the major assets and thus determines the overall quality requirements, the direction and the general set up of the risk-based testing process. \emph{Business} related assets are required for a successful business practice and thus often directly relate to software quality properties like functionality, availability, security and reliability. \emph{Safety} relates to the inviolability of human health and life and thus requires software to be failsafe, robust and resilient. \emph{Security} addresses the resilience of IT systems against threats that jeopardize confidentiality, integrity and availability of digital information and realted services. Finally, \emph{Compliance} relates to assets that are directly derived from rules and regulations be it applicable laws, standards or other forms of governing settlements. Protection of these assets often but not exclusively relate to quality properties like security, reliability and compatibility.

\subsubsection{Quality property} 
\label{sec:quality-property}
A \emph{quality property} is a distinct quality attribute~\cite{ISO_25010} which contributes to the protection of assets and thus, is suject to risk assessment and testing. As stated in~\cite{ISO14971}, risks result from hazards. Hazards related to software-based systems stem from software vulnerabilities and from defects in software functionalities, which are critical to business cases, safety-related aspects, security of systems or applicable rules and regulations. 
 
One needs to test that a software-based system is
\begin{itemize}
    \item functionally suitable, i.e., able to deliver services as requested
	\item reliable, i.e., able to deliver services as specified over a period of time
	\item usable, i.e., satisfies the user expectation
    \item performant and efficient, i.e., able to react appropriately with respect to stated resources and time 
	\item secure, i.e., able to remain protected against accidental or deliberate attacks
	\item resilient, i.e., able to recover timely from unexpected events
	\item safe, i.e., able to operate without harmful states
	\end{itemize}

The quality properties considered determine which testing is appropriate and has to be chosen. We consider \emph{functionality}, \emph{security}, and \emph{reliability} to be the dominant quality properties that are addressed for software. They together form the reliability, availability, safety, security, and resilience of a software-based system and hence constitute the options for the risk drivers in the RBT taxonomy. 

As reported by different computer emergency response teams such as GovCERT-UK, software defects continue to be a major, if not the main source of incidents caused by software-based systems. The quality properties determine the test types and test techniques that are applied in a test process to find software defects or systematically provide belief in the absence of such defects. Functional testing is likewise a major test type in RBT to analyze reliability and safety aspects, see, e.g.,~\cite{Amland2000RBT}. In addition, security testing including penetration testing, fuzz testing and/or randomized testing is key in RBT~\cite{zech2011risk, ETSI_TR101583} to analyze security and resilience aspects. Furthermore, performance and scalability testing focusing on normal load, maximal load, and overload scenarios to analyze availability and resilience aspects, see, e.g.,~\cite{Amland2000RBT}.

\subsubsection{Risk item} 
\label{sec:quality-item-type}
The \emph{risk item} characterizes and determines the elements under evaluation. These risk items are the elements to which risk exposures and tests are assigned~\cite{felderer2013experiences}. Risk items can be of type \emph{test case}~\cite{yoon2011test}, i.e., directly test cases themselves as in regression testing scenarios, \emph{runtime artifact} like deployed services, \emph{functional artifact} like requirements or features, \emph{architectural artifact} like component, or \emph{development artifact} like source code file. The risk item type is determined by the test level. For instance, functional or architectural artifacts are often used for system testing, and generic risks for security testing. In addition, we use the term \emph{artifact} to openly refer to other risk items used in requirements capturing, design, development, testing, deployment, and/or operation and maintenance, which all might relate to the identified risks.

\subsection{Risk assessment}
\label{sec:risk-assessment}
The second differentiating element of RBT approaches is the way risks are being determined. According to~\cite{istqb2012standardGlossary}, \emph{risk assessment} is the process of \emph{identifying} and subsequently \emph{analyzing} the identified risk to determine its level of risk, typically by assigning likelihood and impact ratings. Risk assessment itself has multiple aspects, so that one needs to differentiate further into the \emph{factors} influencing risks, the risk \emph{estimation technique} used to estimate and/or evaluate the risk, the \emph{scale} type that is used to characterize the risk exposure, and the \emph{degree of automation} for risk assessment.

\subsubsection{Factor} 
\label{sec:factors}
The risk factors quantify identified risks~\cite{bai2012risk}. \emph{Risk exposure} is the quantified potential for loss. It is calculated by the likelihood of risk occurrence multiplied by the potential loss, also called the impact. The risk exposure considers typically aspects like liability issues, property loss or damage, and product demand shifts. RBT approaches might also consider the specific aspect of \emph{likelihood} of occurrence, e.g., for test prioritization or selection or the specific aspect of \emph{impact rating} to determine test efforts needed to analyze the countermeasures in the software.

\subsubsection{Estimation technique}
\label{sec:estimation-technique}
The estimation technique determines how the risk exposure is actually estimated and can be \emph{list-based} or \emph{formal model}~\cite{Jorgensen2009EffortEstimation}. The essential difference between formal-model-based and list-based estimation is the quantification step-that is, the final step that transforms the input into the risk estimate. Formal risk estimation models are based on a complex, multi-valued quantification step such as a formula or a test model. On the other hand, list-based estimation methods are based on a simple quantification step-for example, what the expert believes is riskiest. List-based estimation processes range from pure 'gut feelings' to structured, historical data including failure history and checklist-based estimation processes. 

\subsubsection{Scale}
Any risk estimation uses a scale to determine the risk ``level''. This risk scale can be \emph{quantitative} or \emph{qualitative}. Quantitative risk values are numeric and allow computations, qualitative risk values can only be sorted and compared. An often used qualitative scale for risk levels is low, medium, and high~\cite{wendland2012systematic}.

\subsubsection{Degree of automation} 

Risk assessment can be supported by automated methods and tools. For example, risk-oriented metrics can be measured \emph{manually} or \emph{automatically}. The manual measurement is often supported by strict guidelines and the automatic measurement is often performed via static analysis tools. Other examples for automated risk assessment include the derivation of risk exposures from formal risk models, see, for instance,~\cite{FredriksenKGSOD02}.

\subsection{Risk-based testing strategy}
\label{sec:risk-based-test-process}
Based on the risks being determined and characterized, RBT follows the fundamental test process~\cite{istqb2012standardGlossary} or variations thereof. The notion of risk can be used to optimize already existing testing activities by introducing risk-based strategies for prioritization, automation, selection, resource planning etc. Dependent on the approach, nearly all activities and phases in a test process may be impacted by taking a risk-based perspective. This taxonomy aims for highlighting and characterizing the RBT specifics by relating them to the major phases of a normal test process. For the sake of brevity, we have focused on the phases \emph{risk-based test planning}, \emph{risk-based test design \& implementation}, and \emph{risk-based test execution \& evaluation} that are outlined in the following subsections. 

\subsubsection{Risk-based test planning}
\label{sec:risk-based-test}
The main outcome of test planning is a test strategy and a plan that depicts the staffing, the required resources, as well as a schedule for the individual testing activities. \emph{Test planning} establishes or updates the scope, approach, resources, and schedule of intended test activities. Amongst others, \emph{test objectives}, \emph{test techniques}, and \emph{test completion criteria}, which impact risk-based testing~\cite{redmill2005theory}, are determined. 

\paragraph{Test objective \& technique.} \emph{Test objectives \& techniques} are relevant parts of a test strategy. They determine what and how to test a test item. The reason to design or execute a test, i.e., a \emph{test objective}, can be related to the risk item to be tested, to the thread scenarios of a risk item or to the counter measures established to secure that risk item, see also Section~\ref{RBTdesign}. The selection of adequate \emph{test techniques} can be done of basis of the \emph{quality properties} as well as from information related to defects, vulnerabilities and threat scenarios coming from risk assessment. 

\paragraph{Test completion criterion.} Typical exit criteria for testing that are used to report against and to plan when to stop testing, include all tests ran successfully, all issues have been retested and signed off, or all acceptance criteria have been met. Specific RBT-related exit criteria~\cite{Amland2000RBT} add criteria on the residual risk in the product and coverage-related criteria: all risk items, their threat scenarios and/or counter measures being covered. Risk-based metrics are used to quantify different aspects in testing such as the minimum level of testing, extra testing needed because of high number of faults found, the quality of the tests and the test process. They are used to manage the RBT process and optimize it with respect to time, efforts, and quality~\cite{Amland2000RBT}.

\paragraph{Resource planning \& scheduling.} RBT requires focusing the testing activities and efforts based on the risk assessment of the particular product or of the project, in which it is developed. In simple words: if there is high risk, then there will be serious testing. If there is no risk, then there will be rather little testing. For example, products with high complexity, new technologies, many changes, many defects found earlier, developed by personnel with less experiences or lower qualification, or developed along new or renewed development processes may have a higher probability to fail and need to be tested more thoroughly. Within this context, information from risk assessment can be used to roughly identify high-risk areas or features of the system under test (SUT) and thus determine and optimize the respective test effort, the required personnel and their qualification and the scheduling and prioritization of the activities in a test process.

\subsubsection{Risk-based test design \& implementation }
\label{RBTdesign}

\emph{Test design} is the process of transforming test objectives into test cases. This transformation is guided by the coverage criteria, which are used to quantitatively characterize the test cases and often used for exit criteria. Furthermore, the technique of transformation depends on the test types needed to realize a test objective. These test types directly relate to the \emph{quality property} defined in Section \ref{sec:contextual-set-up}. \emph{Test implementation} comprises tasks like preparing test harnesses and test data, providing logging support or writing automated test scripts to enable the automated execution of test cases~\cite{istqb2012standardGlossary}. Risk aspects are especially essential for providing \emph{logging support} and for \emph{test automation}.

\paragraph{Coverage item determination.} RBT uses coverage criteria specific to the risk artifacts and test types specific to the risk drivers on functionality, security, and safety. The classical code-oriented and model-based coverage criteria like path coverage, condition-oriented coverage criteria like modified condition decision coverage, requirements-oriented coverage criteria like requirements or use case coverage are extended with coverage criteria to cover selected or all assets, threat scenarios, and counter measures~\cite{Stallbaum:2008:ATR:1370042.1370057}. While \emph{asset coverage} rather belongs to requirements-oriented coverage~\cite{wendland2012systematic}, \emph{threat scenario \& vulnerability coverage}, and \emph{counter measure coverage} can be addressed by code-oriented, model-based, and/or condition-oriented coverage criteria~\cite{hosseingholizadeh2010source}.

\paragraph{Test or feature prioritization \& selection.} In order to optimize the costs of testing and/or the quality and fault detection capability of testing, techniques for prioritizing, selecting, and minimizing tests as well as combinations thereof have been developed and are widely in use~\cite{Yoo:2012:RTM:2284811.2284813}. In the ranges of intolerable risk and ``As Low As Reasonably Practicable (ALARP)''\footnote{The ALARP principle is typically used for safety-critical, but also for mission-critical systems. It says that the residual risk shall be as low as reasonably practical.} risks, these techniques are used to identify tests for the risk-related test objectives determined before. For example, design-based approaches for test selection~\cite{Briand:2009:ART:1465742.1466092} and coverage-based approaches~\cite{Amland2000RBT} for test prioritization are well-suited for RBT. Dependent on the approach prioritization \& selection can take place during different phases of the test process. A risk-based feature or requirement prioritization \& selection selects the requirements or features to be tested. This activity is usually started during test planning and continued during test design. Test case prioritization \& selection requires existing test specifications or test cases. It is thus either carried out before test implementation to determine the test case to be implemented or in the preparation of test execution or regression testing to determine the optimal test sets to be executed. 

\paragraph{Test case derivation/generation.} Risk assessment often comprises information about threat scenarios, faults, vulnerabilities that can be used to derive the test data, the test actions, probably the expected results and other testing artifacts. Especially when addressing publicly known threat scenarios, these scenarios can be used to directly refer to predefined and reusable test specification fragments i.e., so called test pattern. These test patterns already contain test actions and test data that are directly applicable to either test specification, test implementation or test execution~\cite{Botella2014}. 

\paragraph{Test automation.} Test automation is the use of special software (separate from the software under test) to control the execution of tests and the comparison of actual outcomes with predicted outcomes~\cite{huizinga2007automated}. Experiences from test automation~\cite{graham2012experiences} show possible benefits like improved regression testing or a positive return on investment, but also caveats like high initial investments or difficulties in test maintenance. Risks may therefore be beneficial to guide decisions where and to what degree testing should be automated.  

\subsubsection{Risk-based test execution \& evaluation}
\label{sec:risk-based-test-2}
\emph{Test execution} is the process of running test cases. In this phase, risk-based testing is supported by \emph{monitoring} and \emph{risk metrics measurement}. \emph{Test evaluation} comprises decisions on the basis of exit criteria and logged test results compiled in a test report. In this respect, risks are \emph{mitigated} and may require a \emph{re-assessment}. Furthermore, risks may guide \emph{test exit decisions} and \emph{reporting}.

\paragraph{Monitoring \& risk metrics measurement.} Monitoring is run concurrently with a system under test and supervises, records or analyzes the behavior of the running system~\cite{Radatz1990IEEEStandardGlossary,istqb2012standardGlossary}. Differing from software testing, which actively stimulates the system under test, monitoring only passively observes a running system. For risk-based testing purposes, monitoring enables additional complex analysis, e.g., of the internal state of a system for security testing, as well as tracking the project's progress toward resolving its risks and taking corrective action where appropriate. \emph{Risk metrics measurement} determines risk metrics defined in the test planning phase. A measured risk metric could be the number of observed critical failures for risk items where failure has high impact~\cite{felderer2013using}.

\paragraph{Risk reporting.} Test reports are documents summarizing testing activities and results~\cite{istqb2012standardGlossary} that communicate risks and alternatives requiring a decision. They typically report progress of testing activities against a baseline (such as the original test plan) or test results against exit criteria. In \emph{risk reporting}, assessed risks which are monitored during the test process, are explicitly reported in relation to other test artifacts. Risk reports can be descriptive summarizing relationships of the data or predictive using data and analytical techniques to determine the probable future risk. Typical descriptive risk reporting techniques are risk burn down charts which visualize the development of the overall risk per iteration as well as traffic light reports providing a high level view on risks using colors red for high risks, yellow for medium risks and green for low risks. A typical predictive risk reporting technique is residual risk estimation, for instance, based on software reliability growth models~\cite{goel1985reliability-models}.

\paragraph{Test \& risk re-assessment.} The \emph{re-assessment of risks} after test execution may be planned in the process or triggered by a comparison of test results against the assessed risks. This may reveal deviations between the assessed and the actual risk level and require a re-assessment to adjust them. Test results can explicitly be integrated into a formal risk analysis model~\cite{stallbaum2007employing} or just trigger the re-assessment in an informal way.

\paragraph{Test exit decision.} The \emph{test exit decision} determines if and when to stop testing~\cite{felderer2013experiences}, but may also trigger further risk mitigation measures. This decision may be taken on the basis of a test report matching test results and exit criteria or ad hoc, for instance, solely on the basis of the observed test results.

\paragraph{Risk mitigation.} \emph{Risk mitigation} covers efforts taken to reduce either the likelihood or impact of a risk~\cite{571734}. In the context of risk-based testing, the assessed risks and their relationship to test results and exit criteria (which may be outlined in the test report), may trigger additional measures to reduce either the likelihood or impact of a risk to occur in the field. Such measures may be bug fixing, re-design of test cases or re-execution of test cases.

\section{Classification of Recent Risk-Based Testing Approaches}
\label{sec:class-recent-risk}

In this section, we present four recent risk-based testing approaches, i.e., the RASEN approach (Section~\ref{sec:the-rasen-approach}), the SmartTesting approach (Section~\ref{sec:smarttesting-approach}), risk-based test case prioritization based on the notion of risk exposure (Section~\ref{sec:riskexposure-approach}), as well as risk-based testing of open source software (Section~\ref{sec:rbtoss-approach}), and position them in the risk-based testing taxonomy presented in the previous section.

\subsection {The RASEN approach}
\label{sec:the-rasen-approach}
\subsubsection {Description of the approach}
\begin{figure*}
\begin{center}
  \includegraphics[width=0.7\textwidth]{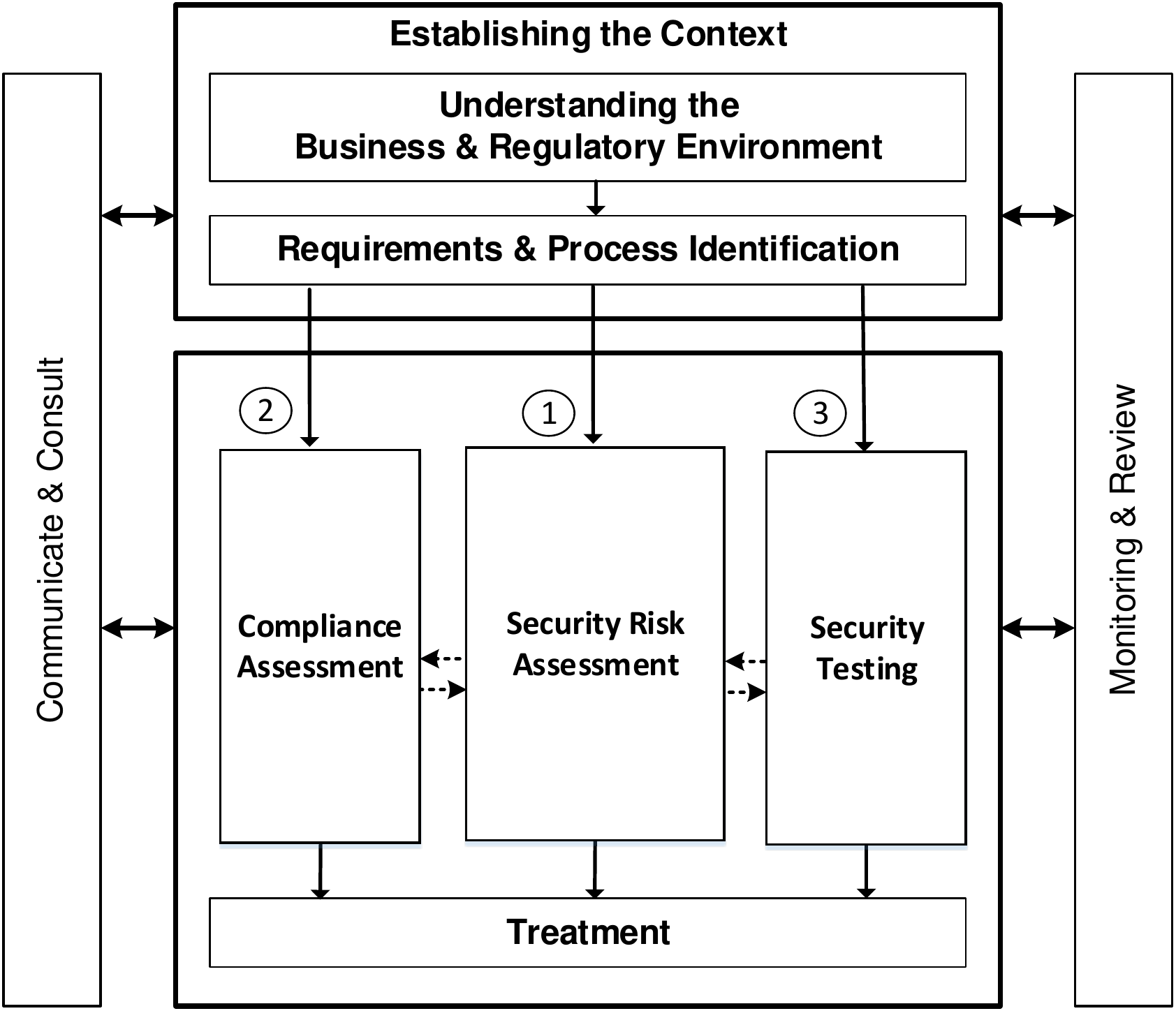}
  \caption{Combining compliance assessment, security risk assessment, and security testing in RASEN}
  \label{fig:rasen-approach}
\end{center}
\end{figure*}

The RASEN project (www.rasen-project.eu) has developed a process for combining compliance assessment, security risk assessment and security testing based on existing standards like ISO-31000 and ISO-29119. The approach is currently extended in the PREVENT project (http://www.prevent-project.org) to cover business driven security risk and compliance management for critical banking infrastructures. Figure~\ref{fig:rasen-approach} shows an overview of the RASEN process. 

The process covers three distinguishable workstreams that each consist of a combination of typical compliance assessment, security risk assessment activities and/or security testing activities emphasizing the interplay and synergies between these former independent assessment approaches.

\begin{enumerate}
\item The test-based security risk assessment workstream starts like a typical risk assessment workstream and uses testing results to guide and improve the risk assessment. Security testing is used to provide feedback on actually existing vulnerabilities that have not been covered during risk assessment or allows risk values to be adjusted on the basis of tangible measurements like test results. Security testing should provide a concise feedback whether the properties of the target under assessment have been really met by the risk assessment. 
\item The risk-based compliance assessment workstream targets the identification and treatment of compliance issues. It relies on security risk assessment results to identify compliance risk and thus systematize the identification of compliance issues. Moreover, legal risk assessment may be used to prioritize the treatment of security issues.
\item The risk-based security testing workstream starts like a typical testing workstream and uses risk assessment results to guide and focus the testing. Such a workstream starts with identifying the areas of risk within the target's business processes and building and prioritizing the testing program around these risks. In this setting risks help focus the testing resources on the areas that are most likely to cause concern or support the selection of test techniques dedicated to already identified threat scenarios. 
\end{enumerate}

According ISO 31000, all workstreams start with a preparatory phase called \emph{Establishing the Context} that includes preparatory activities like understanding the business and regulatory environment as well as the requirements and processes. During this first phase the high-level security objectives are identified and documented and the overall process planning is done. Moreover, the process shows additional support activities like \emph{Communication \& Consult} and \emph{Monitoring and Review} that are meant to set up the management perspective, thus to continuously control, react, and improve all relevant information and results of the process. From a process point of view, these activities are meant to provide the contextual and management related framework. The individual activities covered in these phases might differ in detail dependent on whether the risk assessment or testing activities are the guiding activities. The main phase, namely the \emph{Security Assessment} phase covers the definition of the integrated compliance assessment, risk assessment and a security testing workstreams. 

\paragraph {The risk assessment workstream.} The overall risk assessment workstream is decomposed into the three main activities \emph{Risk Identification}, \emph{Risk Estimation}, and \emph{Risk Evaluation}. RASEN has extended the risk identification and risk estimation activities with security testing activities in order to improve the accuracy and efficiency of the overall workstream.

Risk identification is the process of finding, recognizing and describing risks. This consists og identifying sources of risk (e.g., threats and vulnerabilities), areas of impacts (e.g., the assets), malicious events, their causes and their potential impact on assets. In this context, security testing is used to obtain information that eases and supports the identification of threats and threat scenarios. Appropriate are testing and analysis techniques that yield information about the interfaces and entry points (i.e., the attack-surface) like automated security testing, network discovery, web-crawling, and fuzz testing. 

Following risk identification, risk estimation is the process of expressing the likelihood, intensity, and magnitude of the identified risks. In many cases, the relevant information on potential threats are often imprecise and insufficient, so that estimation often relies on expert judgment only. This, amongst others, might result in a high degree of uncertainty related to the correctness of the estimates. Testing or test-based risk estimation may increase the amount of information on the target of evaluation. Testing might in particular provide feedback regarding the resilience of systems, i.e., it can support the estimation of the likelihood that an attack will be successful if initiated. Information from testing on the presence or absence of potential vulnerabilities have direct impact on the likelihood values of the associated threat scenarios. Similar to test-based risk identification, penetrating testing tools, model-based security testing tools, static and dynamic code analysis tools, and vulnerability scanners are useful to obtain this kind of information.

\paragraph {The compliance assessment workstream.} The risk-based compliance assessment workstream consists of three major steps. The compliance risk identification step provides a systematic and template based approach to identify and select compliance requirements that imply risk. These requirements are transformed into obligations and prohibitions that are the basis for further threat and risk modelling using the CORAS tool. The second step, the compliance risk estimation step is dedicated to understanding and documenting the uncertainty that originates from compliance requirements interpretation. Uncertainty may arise from unclear compliance requirements or from uncertainty about the consequences in case of non-compliance. During compliance risk evaluation compliance requirements are evaluated and prioritized based on their level of risk so that during treatment compliance resources may be allocated efficiently based on their level of risk. In summary, combining security risk assessment and compliance assessment helps prioritizing compliance measures based on risks and helps to identify and deal with compliance requirements that directly imply risk. 

\paragraph {The security testing workstream.} The risk-based security testing workstream is structured like a typical security testing process. It starts with a test planning phase, followed by a test design \& implementation phase and ends with test execution, analysis and summary. The result of the risk assessment, i.e., the identified vulnerabilities, threat scenarios and unwanted incidents, are used to guide the test planning, test identification and may complement requirements engineering results with systematic information concerning the threats and vulnerabilities of a system.

Factors like probabilities and consequences can be additionally used to weight threat scenarios and thus help identifying which threat scenarios are more relevant and thus identifying the ones that need to be treated and tested more carefully. From a process point of view, the interaction between risk assessment and testing could be best described following the phases of a typical testing process. 

\begin{enumerate}
 \item Risk-based security test planning deals with the integration of security risk assessment in the test planning process.
 \item Risk-based security test design and implementation deals with the integration of security risk assessment in the test design and implementation process. 
 \item Risk-based test execution, analysis and summary deals with a risk-based test execution as well as with the systematic analysis and summary of test results. 
\end{enumerate}

\subsubsection{Positioning in the risk-based testing taxonomy.}

\paragraph{Context.} The overall process \cite{ETSI_EG203251, Grossmann2015} is directly derived from ISO-31000 and slightly extended to highlight the integration with security testing and compliance assessment. The approach explicitly addresses \emph{compliance} but also \emph{business} and in a limited way \emph{safety} as major \emph{risk drivers}. It is defined independent from any application domain and independent from the level, target or depth of the security assessment itself. It could be applied to any kind of technical assessment process with the potential to target the full number of \emph{quality properties} that are defined in Section \ref{sec:quality-property}. Moreover, it addresses legal and compliance issues related to data protection and security regulations. Looking at risk-based security testing, the approach emphasizes executable \emph{risk items}, i.e., \emph{runtime artifacts}. Considering risk-based compliance assessment, the approach also addresses the other risk items mentioned in the taxonomy.  

\paragraph{Risk assessment.} The test-based risk assessment workstream uses test results as explicit input to various activities of the risk assessment. Risk assessment in RASEN has been carried out on basis of the CORAS method and language. Thus, risk estimation is based on \emph{formal} models that support the definition of \emph{likelihood} values for events and \emph{impact} values to describe the effect of incidents on assets. Both, likelihood and impact values are used to calculate the overall \emph{risk exposure} for unwanted incidents, i.e., the events that directly harm assets. CORAS is flexible with respect to the calculation scheme as well as to the scale for defining risk factors. It generally supports values with \emph{qualitative scale} as well as with \emph{quantitative scale}.

\paragraph{Risk-based test strategy.} Security is a non-functional property and thus requires dedicated information that addresses the (security) context of the system. While functional testing is more or less guided directly by the system specification (i.e., features, requirements, architecture), security testing often is not. The RASEN approach to \emph{risk-based security test planning} especially addresses the risk-based selection of \emph{test objectives \& test techniques} as well as risk-based \emph{resource planing \& and scheduling}. Security risk assessment is serving this purpose and can be used to roughly identify high-risk areas or features of the system under test (SUT) and thus determine and optimize the respective test effort. Moreover, a first assessment of the identified vulnerabilities and threat scenarios may help to select test strategies and techniques that are dedicated to deal with the most critical security risks. Considering \emph{security test design \& implementation}, especially the selection and prioritization of the feature to test, the concrete tests design and the determination of \emph{test coverage items} are critical. A recourse to security risks, potential threat scenarios and potential vulnerabilities provide a good guidance to improve \emph{item prioritization \& selection}. Security risk related information support the selection of features and test conditions that require testing. It helps in identifying which coverage items should be covered in which depth and how individual test cases and test procedures should look. The RASEN approach to risk-based security test design \& implementation uses information on expected threats and potential vulnerabilities to systematically determine and identify coverage items (besides others \emph{asset coverage}, \emph{threat scenario \& vulnerabilities coverage} ), test conditions (testable aspects of a system) and test purposes. Moreover, the security risk assessment provides quantitative estimations on the risks, i.e., the product of frequencies or probabilities and estimated consequences. This information is used to select and prioritize either the test conditions or the actual tests when they are assembled into test sets. Risks as well as their probabilities and consequence values are used to set priorities for the test selection, test case generation as well as for the order of test execution expressed by risk-optimized test procedures. Risk-based test execution allows the prioritization of already existing test cases, test sets or test procedures during regression testing. \emph{Risk-based security test evaluation} aims for improving \emph{risk reporting} and the \emph{test exit decision} by introducing the notion of risk coverage and remaining risks on basis of the intermediate test results as well as on basis of the errors, vulnerabilities or flaws that have been found during testing. In summary we have identified the three activities that are supported through results from security risk assessment.

\subsection{The SmartTesting Approach}
\label{sec:smarttesting-approach}

\subsubsection {Description of the approach}

\begin{figure*}[htb]
\begin{center}
  \includegraphics[width=.7\textwidth]{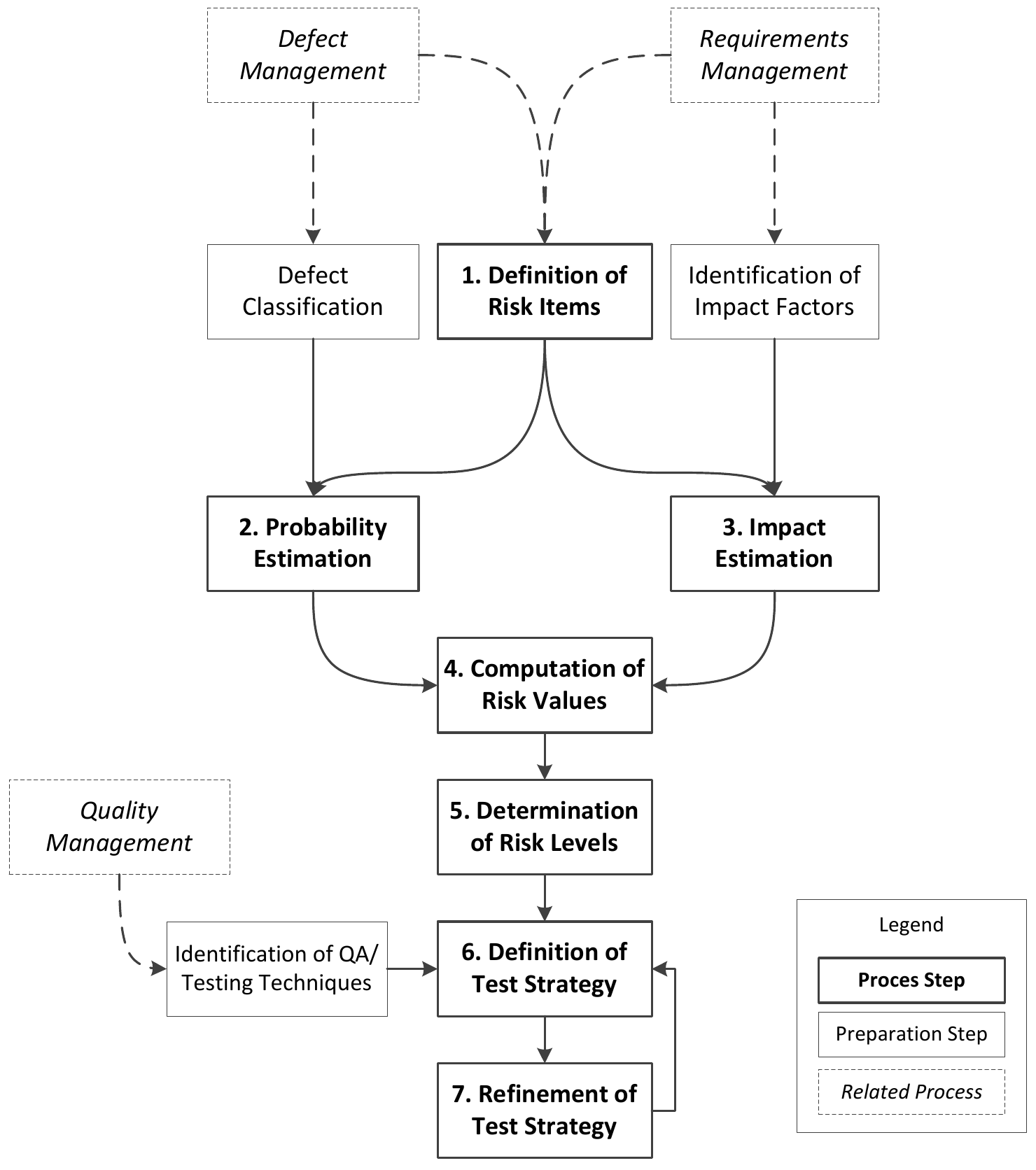}
  \caption{SmartTesting Process~\cite{ramler2015process}}
  \label{fig:smarttesting-process}
\end{center}
\end{figure*}

Figure~\ref{fig:smarttesting-process} provides an overview of the overall process. It consists of different steps, which are either directly related to the risk-based test strategy development (shown in bold font) or which are used to establish the preconditions (shown in normal font) for the process by linking test strategy development to the related processes (drawn with dashed lines) of defect management, requirements management and quality management. The different steps are described in detail in the following subsections.

\paragraph{Definition of risk items.} In a first step, the risk items are identified and defined. The risk items are the basic elements of a software product that can be associated with risks. Risk items are typically derived from the functional structure of the software system, but they can also represent non-functional aspects or system properties. In the context of testing it should be taken into account that the risk items need to be mapped to test objects~\cite{istqb2012standardGlossary}, i.e., testable objects such as sub-systems, features, components, modules or functional as well as non-functional requirements. 

\paragraph{Probability estimation.} In this step the probability values (for which an appropriate scale has to be defined) are estimated for each risk item. In the context of testing the probability value expresses the likelihood of defectiveness of a risk item, i.e., the likelihood that a fault exists in a specific product component due to an error in a previous development phase that may lead to a failure. There are several ways to estimate or predict the likelihood of a component's defectiveness. Most of these approaches rely on historical defect data collected from previous releases or related projects. Therefore, defect prediction approaches are well suited to support probability estimation~\cite{ramler2016defectsrbt}.

\paragraph{Impact estimation.} In this step the impact values are estimated for each risk item. The impact values express the consequences of risk items being defective, i.e., the negative effect that a defect in a specific component has on the user or customer and, ultimately, on the company's business success. The impact is often associated with the cost of failures. The impact is closely related to the expected value of the components for the user or customer. The value is usually determined in requirements engineering when eliciting and prioritizing the system's requirements. Thus, requirements management may be identified as main source of data for impact estimation. 

\paragraph{Computation of risk values.} In this step risk values are computed from the estimated probability and impact values. Risk values can be computed according to the definition of risk as $R = P \times I$ where P is the probability value and I is the impact value. Aggregating the available information to a single risk value per risk item allows the prioritization of the risk items according to their associated risk values or ranks. Furthermore, the computed risk values can be used to group risk items, for example, according high, medium and low risk. Nevertheless, for identifying risk levels it is recommended to consider probability and impact as two separate dimensions of risk. 

\paragraph{Determination of risk levels.} In this step the spectrum of risk values is partitioned into risk levels. Risk levels are a further level of aggregation. The purpose of distinguishing different risk levels is to define classes of risks such that all risk items associated to a particular class are considered equally risky. As a consequence, all risk items of the same class are subject to the same intensity of quality assurance and test measures.

\paragraph{Definition of test strategy.} In this step the test strategy is defined on the basis of the different risk levels. For each risk level the test strategy describes how testing is organized and performed. Distinguishing different levels allows testing to be performed with differing levels of rigor in order to adequately address the expected risks. This can either be by achieved by applying specific testing techniques (e.g., unit testing, use case testing, beta testing, reviews) or by applying these techniques with more or less intensity according to different coverage criteria (e.g., unit testing at the level of 100\% branch coverage or use case testing for basic flows and/or alternative flows).

\paragraph{Refinement of test strategy.} In the last step the test strategy is refined to match the characteristics of the individual components of the software system (i.e., risk items). The testing techniques and criteria that have been specified in the testing strategy for a particular risk level can be directly mapped to the components associated with that risk level. However, the test strategy is usually rather generic. It does not describe the technical and organizational details that are necessary for applying the specified techniques to a concrete software component. For each component, thus, a test approach has to be developed that clarifies how the test strategy should be implemented.

\subsubsection{Positioning in the risk-based testing taxonomy}

\paragraph{Context.} SmartTesting provides a lightweight process for development and refinement of a risk-based test strategy. It does not explicitly address risk drivers, but - as in every risk-based testing process - implicitly it is assumed that a risk driver and a quality property to be improved are available. The risk drivers of the broad range of companies involved in the accompanying study~\cite{ramler2015process} cover all types, i.e., business, safety and compliance. Also different quality properties of interest are covered, mainly as impact factors. For instance, in the involved companies considered performance and security besides functionality as impact factors. 

\paragraph{Risk assessment.} SmartTesting explicitly contains a step to define risk items, which can in principle be of any type from the taxonomy. In the case companies, risk items were typically derived from the system's component structure. Via the process step computation of risk values, SmartTesting explicitly considers \emph{risk exposure}, which is \emph{qualitatively} estimated by a mapping of risk values to risk levels in the process step determination of risk levels. The risk value itself is measured based on a \emph{formal model} in the process step computation of risk values, which combines values from probability and impact estimation. Probability estimation takes defect data into account, and impact estimation is based on impact factors, which are typically \emph{assessed manually}. 

\paragraph{Risk-based test strategy.} The process steps definition and refinement of the test strategy comprises \emph{risk-based test planning} resulting in the assignment of concrete techniques, resource planning and scheduling, prioritization and selection strategies, metrics as well as exit criteria to the risk levels and further to particular risk items.

\subsection{Risk-based test case prioritization based on the notion of risk exposure}
\label{sec:riskexposure-approach}

\subsubsection{Description of the approach}
Choi et. al. present different test case prioritization strategies based on the notion of \emph{risk exposure}. In \cite{yoon2011test}, test case prioritization is described as an activity with the aim ``to find the most important defects as early as possible against the lowest costs'' \cite{redmill2005theory}. Choi et. al. claim that their risk-based approach to test case prioritization performs well against this background. They empirically evaluate their approach in a setting where various versions of a Traffic Conflict Avoidance System (TCAS) are tested and show how their approach performs well compared to the prioritization approach of others. In \cite{Hettiarachchi20161} the approach is extended using an improved prioritization algorithm and towards an automated risk estimation process using fuzzy expert systems. A fuzzy expert system is an expert system that uses fuzzy logic instead of Boolean logic to reason about data. Conducting risk estimation with this kind of expert system, Choi et al. aim to replace the human actor during risk estimation and thus avoid subjective estimation results. The second approach has been evaluated by prioritizing test cases for two software products, the electronic health record software \emph{iTrust}, an open source product, and an industrial software application called \emph{Capstone}. 

\subsubsection{Positioning in the risk-based testing taxonomy}

\paragraph{Context.} Both approaches do not explicitly mention one of the risk drivers from Section \ref{sec:risk-driver} nor do they provide exhaustive information on the addressed quality properties. However, in \cite{yoon2011test} the authors evaluate their approach in context of a safety critical application. Moreover, the authors emphasize that they refer to risks that are identified and measured during the product risk assessment phase. Such a phase is typically prescribed for safety critical systems. Both facts indicate that \emph{safety} seems to be the major\emph{risk driver} and the safety relevant attributes like \emph{functionality},\emph{ reliability} and \emph{performance} are the major quality properties that are addressed by testing. In contrast, the evaluation in \cite{Hettiarachchi20161} is carried out with \emph{business} critical software considering quality properties like \emph{functionality} and \emph{security}. Both approaches have in common that they do not address \emph{compliance} as a risk driver.

\paragraph{Risk assessment.} The risk assessment process for both approaches aim for calculating \emph{risk exposure}. The authors define risk exposure as a value with \emph{quantitative} scale that express the magnitude of a given risk. While in \cite{yoon2011test} the authors explicitly state that they are intentionally not using their own testing related equivalent for expressing risk exposure but directly refer to risk values coming from a pre-existing risk assessment, risk estimation in \cite{Hettiarachchi20161} is done automatically and tailored towards testing. The authors calculate risks on basis of a number of indicators that are harvested from \emph{development artifacts} like requirements. They use properties like requirements modification status and frequency as well as requirements complexity and size to determine the risk likelihood and risk impact for each requirement. In addition indicators on potential security threats are used to address and consider the notion of \emph{security}. In contrast to \cite{yoon2011test}, \cite{Hettiarachchi20161} addresses the \emph{automation} of the risk estimation process using an expert system that is able to aggregate the risk indicators and thus to automatically compute the overall risk exposure. While \cite{yoon2011test} does not explicitly state whether the initial risk assessment relies on formal models or not, the approach in \cite{Hettiarachchi20161} is completely \emph{formal}. However, since \cite{yoon2011test} refers to safety critical systems, we can assume that the assessment is not just a list-based assessment.

\paragraph{Risk-based test strategy.} With respect to testing, both approaches aim for an efficient \emph{test prioritization \& selection} algorithm. Thus, they are mainly applicable in situations where test cases or at least test case specifications are already available. This addresses first of all regression testing but as well decision problems during test management, e.g., when test cases are already specified and the prioritization of test implementation efforts is required. 

To obtain an efficient test prioritization strategy, both approaches aim for deriving risk related weights for individual test cases. In \cite{yoon2011test}, the authors propose two different strategies. The first strategy aims for simple \emph{risk coverage}. Test cases that cover a given risk, obtain a weight that directly relates to the risk exposure for that risk. If a test case covers multiple risks, the risk exposure values are summed. The second strategy additionally tries to consider the fault revealing capabilities of the test cases. Thus, the risk related weight for a test case is calculated by means of the risk exposure for a given risk correlated with the number of risk-related faults that are detectable by that test case, so that test cases with a higher fault revealing capabilities are rated higher. Fault revealing capabilities of test cases are derived through mutation analysis, i.e., this strategy requires that the test cases already exist and that they are executable. 
 
In \cite{Hettiarachchi20161}, test cases are prioritized on basis of their relationship to risk-rated requirements. Risk rating for requirements is determined by an automated risk rating conducted by the fuzzy expert system and an additional analysis of fault classes and their relation to the individual requirements. In short, a fault class is considered to have more impact if it relates to requirements with a higher risk exposure. In addition, a fault of given fault class is considered to occur more often if that fault class relates to a larger number of requirements. Both values determine the overall risk rating for the individual requirements and thus provide the prioritization criteria for requirements. Test cases finally are ordered by means of their relationship to the prioritized requirements. During the evaluation of the approach, the authors obtained the relationship between test cases and requirements from existing traceability information. 

While both approaches provide strong support for \emph{risk-based item selection}, they do not support other activities during \emph{risk-based test design \& implementation} nor do they establish dedicated activities in the area of \emph{risk-based test execution \& evaluation}.

\subsection{Risk-based testing of Open Source Software}
\label{sec:rbtoss-approach}

\subsubsection{Description of the approach}

Yahav et al.~\cite{yahav2014risk,yahav2014data} provide an approach to risk-based testing of open source software (OSS) to select and schedule dynamic testing based on software risk analysis. Risk levels of open source components or projects are computed based on communication between developers and users in the open source software community. Communication channels usually include mail, chats, blogs and repositories of bugs and fixes. The provided data-driven testing approach therefore builds on three repositories, i.e., a social repository which stores the social network data from the mined OSS community, a bug repository which links the community behavior and OSS quality, as well as a test repository which traces test (scripts) to OSS projects. As a preprocessing step, OSS community analytics is performed to construct a social network of communication between developers and users. In a concrete case study~\cite{yahav2014data}, the approach predicts the expected number of defects for a specific project with logistic regression based on the email communication and the time since the last bug. 

\subsubsection{Positioning in the risk-based testing taxonomy}

\paragraph{Context.}
The approach does not explicitly mention one of the risk drivers from Section~\ref{sec:risk-driver} nor of the quality properties. However, the authors state that the purpose for risk-based testing is the experienced significant failures in product quality, timeliness and delivery cost when adapting OSS components in commercial software packages~\cite{yahav2014risk}. Therefore, risk drivers may be \emph{business} to guarantee the success of a system, where the tested OSS component is integrated, or even safety, if the OSS component would be integrated into a \emph{safety}-critical system. Due to the testing context, i.e., selection or prioritization of available tests of OSS components, the main quality property is supposed to be \emph{functionality}. The risk item type are OSS components (developed in OSS projects), i.e., \emph{architectural artifacts}.

\paragraph{Risk Assessment.}
The risk assessment approach quantifies the \emph{likelihood}. For this purpose, a \emph{formal model} is created to predict the number of bugs based on the communication in different communities and the time since the last bug. The scale is therefore \emph{quantitative} as the approach tries to predict the actual number of bugs. The approach implements an \emph{automatic assessment} as it uses monitors to automatically store data in repositories and then applies machine learning approaches, i.e., logistic regression, to predict the risk level.

\paragraph{Risk-based test strategy.}
The approach explicitly supports risk-based test planning in terms of \emph{test prioritization \& selection} and \emph{resource planning \& scheduling}. The approach mainly addresses the allocation of available test scripts for dynamic testing and highlights that exhaustive testing is infeasible and that therefore selective testing techniques are needed to allocate test resources to the most critical components. Therefore, risk-based test design is not explicitly addressed. As \emph{risk metrics} the number of communication metrics as well as the time since the last defect are computed. The approach uses specific \emph{logging support} to log and trace community and defect data. For \emph{risk reporting} confusion matrices are used, which contrast the actual and predicted number of defects.

\section{Summary}
\label{sec:summary}

In this chapter, we presented a taxonomy of risk-based testing. It is aligned with the consideration of risks in all phases of the test process and consists of three top-level classes, i.e., contextual set up, risk assessment, and risk-based test strategy. The contextual set up is defined by risk drivers, quality properties and risk items. Risk assessment comprises the subclasses factors, estimation technique, scale, and degree of automation. The risk-based test strategy then takes the assessed risks into account to guide test planning, test design \& implementation, as well as test execution \& evaluation. The taxonomy provides a framework to understand, categorize, assess and compare risk-based testing approaches to support their selection and tailoring for specific purposes. To demonstrate its application and alignment with available risk-based testing approaches, we positioned four recent risk-based testing approaches, i.e., the RASEN approach, the SmartTesting approach, risk-based test case prioritization based on the notion of risk exposure as well as risk-based testing of Open Source Software, in the taxonomy.

\bibliographystyle{wileynum}

\bibliography{references}

\end{document}